\documentclass[pra,aps,showpacs,showkeys,groupedaddress,superscriptaddress,twocolumn]{revtex4-1}

\usepackage[utf8x]{inputenc}
\usepackage{color}
\usepackage{bbm} 
\usepackage{amsfonts,amsmath,amssymb,stmaryrd}
\usepackage{graphicx}
\usepackage{subfigure}  % use for side-by-side figures
\usepackage{hyperref}
\usepackage{epsfig}
\usepackage{mathrsfs}
\usepackage{verbatim}
\usepackage{centernot}
\usepackage[normalem]{ulem}

\newcommand{\tr}{\text{Tr}}

\renewcommand{\a}{\hat{a}}

\newcommand{\ad}{\hat{a}^\dagger}
\newcommand{\bd}{\hat{b}^\dagger}
\renewcommand{\b}{\hat{b}}

\newcommand{\im}{\text{Im}}
\newcommand{\matr}[1]{\mathbf{#1}}
\newcommand{\id}{1\!\!1}

\usepackage{array}

\usepackage{bm}	% bold in math mode
\renewcommand{\vec}[1]{\bm{#1}}

\begin{document}

%%%%%%%%%%%%%%%%%%%%%%%%%%%%%%%%%%%%%%%%%%%%%%%%%
\title{Absence of topology in Gaussian mixed states of bosons}
%%%%%%%%%%%%%%%%%%%%%%%%%%%%%%%%%%%%%%%%%%%%%%%%%

\author{Christopher D. Mink}
\affiliation{Department of Physics and Research Center OPTIMAS, University of Kaiserslautern, Germany}
\author{Michael Fleischhauer}
\affiliation{Department of Physics and Research Center OPTIMAS, University of Kaiserslautern, Germany}
\author{Razmik Unanyan}
\affiliation{Department of Physics and Research Center OPTIMAS, University of Kaiserslautern, Germany}

%%%%%%%%%%%%%%%%%%%%%%%%%%%%%%%%%%%%%%%%%%%%%%%%%

\begin{abstract}
In a recent paper [Bardyn \emph{et al.} Phys. Rev. X \textbf{8}, 011035 (2018)], it was shown that the generalization of the many-body polarization to mixed states can be used to construct a topological invariant which is also applicable to finite-temperature and non-equilibrium Gaussian states of lattice fermions. 
The many-body polarization defines an ensemble geometric phase (EGP) which is identical to the Zak phase of a fictitious Hamiltonian, whose symmetries determine the topological classification.
Here we show that in the case of Gaussian states of \emph{bosons} the corresponding topological invariant is always trivial.
This also applies to finite-temperature states of bosons in lattices with a topologically non-trivial band-structure.
As a consequence there is no quantized topological charge pumping for translational invariant bulk states of non-interacting bosons.
\end{abstract}

\pacs{03.65.Vf, 03.65.Yz}

\date{\today}
\maketitle

%%%%%%%%%%%%%%%%%%%%%%%%%%%%%%
\section{introduction}
%%%%%%%%%%%%%%%%%%%%%%%%%%%%%%

Topological states of matter have fascinated physicists for many decades as they can give rise to interesting phenomena such as protected edge states and edge currents \cite{Hatsugai-PRL-1993}, quantized bulk transport in insulating states \cite{Klitzing-PRL-1980,TKNN-PRL-1982,Thouless-PRB-1983,Tsui-PRL-1982,Niu-JPhysA-1984,Nakajima-NatPhys-2016} and exotic elementary excitations \cite{Laughlin-PRL-1983,Arovas-PRL-1984,Nayak-RMP-2008}.
Recently, several attempts were made to generalize the concept of topology to finite-temperatures and to non-equilibrium steady states of non-interacting fermion systems \cite{Avron-NJP-2011,Bardyn-NJP-2013,Viyuela-PRL-2014,Huang-PRL-2014,Viyuela-PRL-2014b,Nieuwenburg-PRB-2014,Linzner-PRB-2016,Bardyn-PRX-2018}.
This has been done for fundamental reasons and because of the intrinsic robustness of steady states of driven, dissipative systems.
Integer quantized topological invariants such as the winding of the Berry or Zak phase \cite{Berry-1984,Wilczek-PRL-1984,Zak-PRL-1989,Xiao-RMP-2010} of a one-dimensional band hamiltonian under cyclic parameter variations or the Chern number associated with two-dimensional band structures attain physical significance e.g.~due to the quantization of physical observables 
in insulating states.
Famous examples for this are the charge transport in a Thouless pump \cite{Thouless-PRB-1983,Rice-Mele-PRL-1982,Nakajima-NatPhys-2016} or the Hall conductivity in Chern insulators \cite{Klitzing-PRL-1980,TKNN-PRL-1982,Tsui-PRL-1982,Laughlin-PRL-1983}.  For finite temperatures or under non-equilibrium conditions these quantities are no longer quantized \cite{Wang-PRL-2013}. Furthermore, defining single-particle invariants becomes difficult as the system is in general in a mixed state.
While for one-dimensional systems generalizations of geometric phases to
density matrices based on the Uhlmann construction \cite{Uhlmann-Rep-Math-Phys-1986} can be used \cite{Viyuela-PRL-2014,Huang-PRL-2014}, their application to higher dimensions \cite{Viyuela-PRL-2014b} is faced with difficulties \cite{Budich-Diehl-PRB-2015}.

In a recent paper \cite{Bardyn-PRX-2018}, it was shown that the winding of the many-body polarization introduced by Resta \cite{Resta-PRL-1998} upon a closed path in parameter space is an alternative and useful many-body topological invariant for Gaussian states of fermions.
The polarization of a non-degenerate ground-state $\vert\psi\rangle$ corresponding to a filled band of a lattice Hamiltonian with periodic boundary conditions
is the phase (in units of $2\pi$) induced by a momentum shift $\hat T$ 
\begin{equation}
	P = \frac{1}{2\pi} \textrm{Im} \log \bigl\langle\psi\bigr\vert \hat T \bigl\vert \psi\bigr\rangle.\label{eq:Resta}
\end{equation}
$\hat T$ shifts the lattice momentum $p_k=2\pi k/L$ of all particles by one unit  $\hat T^{-1} \hat c_{\alpha,k} \hat T = \hat c_{\alpha,k+1}$, where $L$ is the number of unit cells
and $\alpha$ a band index. As shown by King-Smith and Vanderbilt \cite{King-Smith-PRB-1993}, expression (\ref{eq:Resta}) for a filled Bloch band is identical to the geometric Zak phase $\phi_\textrm{Zak}$ of this band. The amplitude of $z=\langle\psi\vert \hat T\vert\psi\rangle$, called polarization amplitude, has been used as
indicator for particle localization \cite{Resta-PRL-1998,Resta-Sorella-PRL-1999,Aligia-PRL-1999}. For an insulating many-body state $\vert z\vert$ remains
finite in the thermodynamic limit of infinite  particle number $N\to \infty$, while it vanishes in a 
gapless state \cite{Nakamura-PRB-2002,Kobayashi-PRB-2018}.
 
$P$ can straightforwardly be generalized to mixed states $\rho$ and defines the ensemble geometric phase (EGP) $\phi_\textrm{EGP}$: 
\begin{equation}
	\phi_\textrm{EGP} = \textrm{Im} \log \textrm{Tr}\bigl\{ \rho \hat T\bigr\}.
\end{equation}
Since mixed states are in general not gapped, $\vert \textrm{Tr}\{ \rho \hat T\}\vert$ is expected to vanish in the thermodynamic limit.
However, $\phi_\textrm{EGP}$ remains well defined and meaningful for arbitrarily large but finite systems \cite{Bardyn-PRX-2018} as long as the so-called purity gap of $\rho$ does not close. Furthermore as shown in \cite{Bardyn-PRX-2018} the EGP of a Gaussian density matrix is reduced to the ground-state Zak phase of a \textit{fictitious} Hamiltonian in the thermodynamic limit $L\to \infty$.
The symmetries of this fictitious Hamiltonian determine the topological classification \cite{Bardyn-NJP-2013} following the scheme of Altland and Zirnbauer \cite{Altland-PRB-1997,Schnyder-PRB-2008,Ryu-NJPhys-2010}. A phase transition between different topological phases occurs when the gap of the fictitious Hamiltonian closes for any finite system, i.e.~when $\vert\textrm{Tr}\{\rho \hat T\} \vert=0$.
The many-body polarization is a measurable physical quantity \cite{Bardyn-PRX-2018} and its quantized winding has direct physical consequences. E.g.~it can induce quantized transport in an auxiliary system weakly coupled to a finite-temperature or non-equilibrium system \cite{Wawer-in-prep}. 
It should be noted, however, that due to the absence of a many-body gap, there is in general no adiabatic following in time and the notion of adiabaticity has to be adapted \cite{Bardyn-PRX-2018}.
 
Since the gapfulness of the many-body state is no longer given at finite temperatures, the question arises if the fermionic character of particles is of any relevance and if bosonic Gaussian systems can show non-trivial topological properties as well.
In the present paper we show rigorously that topological invariants based on the many-body polarization are always trivial for Gaussian states of bosons. As a consequence there is e.g.~no 
protected quantized charge pump for bosons under periodic, adiabatic variations of system parameters. 

%%%%%%%%%%%%%%%%%%%%%%%%%%%%%%
\section{the bosonic Rice-Mele model}
%%%%%%%%%%%%%%%%%%%%%%%%%%%%%%

Bloch Hamiltonians with a topologically non-trivial band structure can lead to 
non-trivial many-body invariants of non-interacting fermions, if all single-particle states of the corresponding band(s) are filled.
In such states the many-body polarization can show e.g.~a non-trivial winding under cyclic parameter variations.
Surprisingly, the latter property survives at finite temperatures, i.e.~even if the considered band is no longer fully occupied. Therefore one may ask if the many-body polarization can also show non-trivial behavior in the case  of non-interacting \emph{bosons}? 

\begin{figure}[t]
	\includegraphics[width=\linewidth]{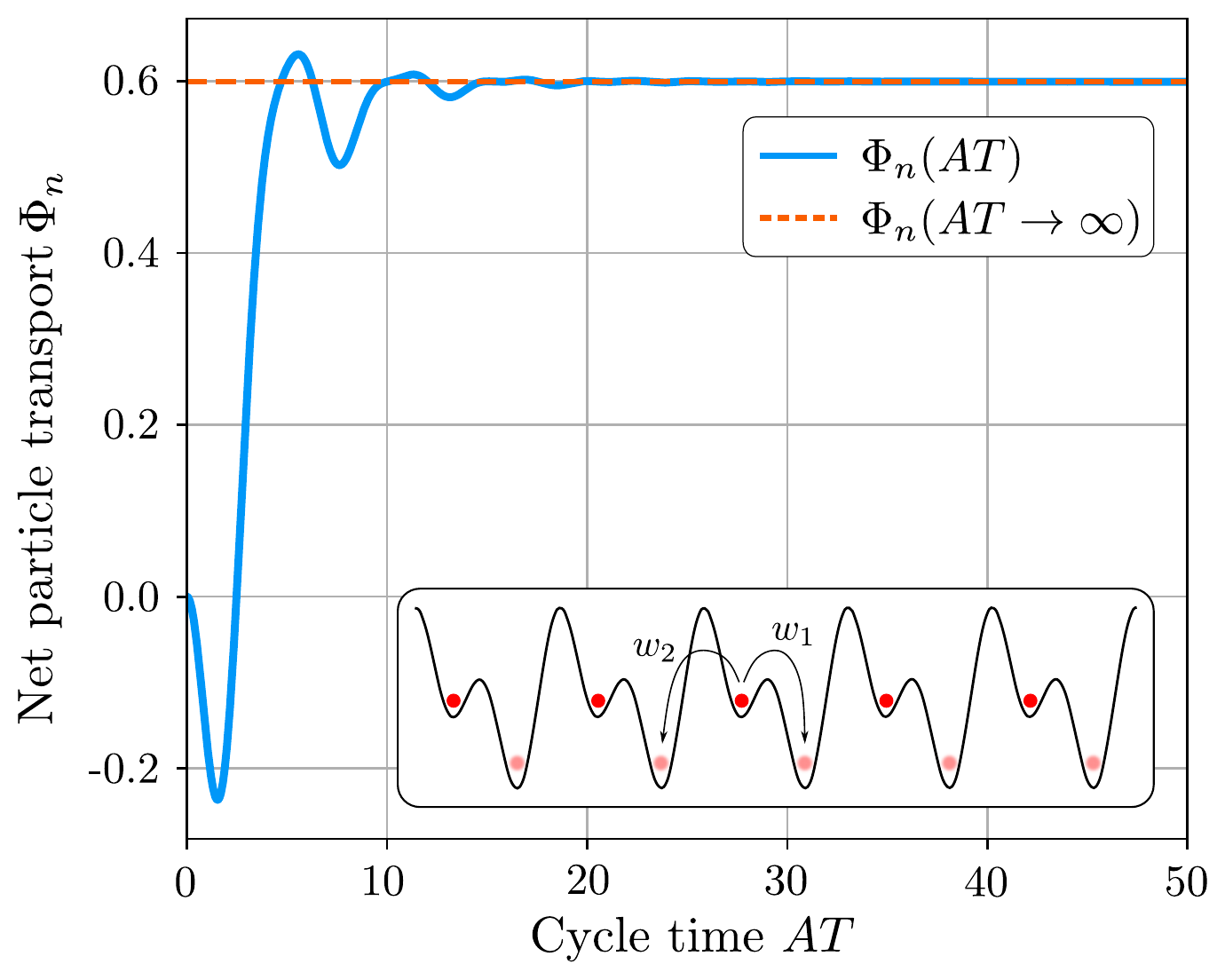}
	\caption{(Color online) Net particle transport as function of the rescaled cycle time $AT$.
		\textit{insert:} Bosonic analogue of Rice-Mele model.
		Non-interacting bosons hop between neighboring lattice sites with alternating hopping rates $w_1$ and $w_2$.  
		The onsite energies are shifted by $\pm \Delta$ in an alternating fashion.}
	\label{fig:flux}
\end{figure}

To illustrate what happens in such a case let us consider one of the simplest 1D lattice models with single-particle topological properties, the Rice-Mele model (RMM) \cite{Rice-Mele-PRL-1982}.
It has a unit cell consisting of two lattice sites with different on-site energies $\pm \Delta$ and describes the hopping of particles with alternating hopping amplitudes $w_{1/2}$ (see insert of Fig.~\ref{fig:flux}).
The Hamiltonian reads
\begin{eqnarray}
	H &=& - w_1\sum_j \ad_j \b_j -w_2 \sum_j \ad_{j+1} \b_j + h.a.\nonumber\\
	&&  + \Delta \sum_j(\ad_j \a_j - \bd_j \b_j) 
	\label{Hamiltonian},
\end{eqnarray}
where $\a_j,\b_j$ are particle annihilation operators at the two sites of the $j$th unit cell and we assume periodic boundary conditions. 
This model is well-known to have a 
% topologically
non-trivial winding of the Zak-phase \cite{Zak-PRL-1989}
\begin{equation}
\phi_\textrm{Zak} = \int_\textrm{BZ} dk  \, \langle u_n(k) \vert \partial_k \vert u_n(k)\rangle
\end{equation}
of anyone of the two subbands $n=1,2$ upon cyclic variations of the parameters $\Delta,w_1 - w_2$ encircling the origin $(\Delta=0,w_1=w_2)$ where the band gap closes. 
Here $\vert u_n(k)\rangle$ are the single-particle Bloch states of the $n$th band at lattice momentum $k\cdot 2\pi/L$.
Performing such a loop adiabatically, one can induce bulk transport if one subband is filled with fermions. At the same time also the many-body polarization 
shows a non-trivial winding which, as shown by King-Smith and Vanderbilt, is strictly connected to the winding of $\phi_\textrm{Zak}$ \cite{King-Smith-PRB-1993}.

Let us now consider the bosonic analogue of the RMM. If initially only one
unit cell is occupied, the center of mass of the wavepacket moves by exactly
one unit cell after a full cycle. This is because this particular initial state
has equal amplitudes in all momentum eigenmodes of the band. The situation is
very different however, when we consider a translationally invariant, periodic
system, where the many-body state returns to itself after a full cycle modulo
a phase factor.
%%%%%%%%%%%%%%%%%%%%%%%%%%%%%%%%%%
%%%%%%%%%%%%%%%%%%%%%%%%%%%%%%%%%%%
Due to translational invariance the Hamiltonian factorizes in momentum modes $\hat a_k,\hat b_k$.
\begin{equation}
H= \sum_k \left(
\hat a_k^\dagger, \hat b_k^\dagger \right)
\mathbf{h}_k(t) 
\left(
\begin{array}{c} 
\hat a_k \\ \hat b_k 
\end{array}\right)
\end{equation}
where $\mathbf{h}_{k}(t)  =\vec{Q}_{k}\left(  t\right)  \cdot\vec{\sigma}$
is a $2\times2$ matrix describing a spin-$\frac{1}{2}$ particle in a magnetic field.
\begin{equation}
\vec{Q}_{k}\left(  t\right) =
\begin{pmatrix}
	w_{1}\left(  t\right)  +w_{2}\left(
	t\right)  \cos\left(  \frac{2\pi k}{L}\right)\\
	w_{2}\left(  t\right)
	\sin\left(  \frac{2\pi k}{L} \right)\\
	\Delta\left(  t\right)
\end{pmatrix}
.\label{effmagnetic}%
\end{equation}
The spectrum of $\mathbf{h}_{k}\left(  t\right)  $ has two bands $\epsilon_{\pm}\left(
k,t\right)  =\pm \epsilon_{k}\left(  t\right)$, where $
\epsilon_{k}\left(  t\right)  =\left[\Delta^{2}\left(  t\right)  +\left\vert
	w_{1}\left(  t\right)  +w_{2}\left(  t\right)  \exp\left(  \frac{2i\pi k}%
	{L}\right)  \right\vert ^{2}\right]^{1/2}$.
The system is assumed to start its evolution at $t=0$, initially being in a
(multi-mode) coherent state. Since the Hamiltonian is quadratic the state remains a coherent state at all times. 
Specifically we consider the initial state
\begin{align}
	\left\vert \Psi\left(  0\right)
	\right\rangle =\underset{\text{Cell }1}{\underbrace{\left\vert \alpha\right\rangle
			\left\vert \beta \right\rangle }}\otimes..\otimes\underset{\text{Cell
		}L}{\underbrace{\left\vert \alpha\right\rangle \left\vert \beta\right\rangle }},\label{eq:coherent}
\end{align}
with $\vert \alpha\vert^2+\vert\beta\vert^2=1$, 
i.e. all cells are occupied equally with average occupation of one per unit cell. 
We note that for coherent states the particle number does not have a well
defined value. Furthermore in contrast to the case of non-interacting fermions this state corresponds
to an initial occupation of only the $k=0$ mode. 
Since the bosons are non-interacting, all
initially empty modes ($k\neq0$) remain empty during the time evolution. Thus, to describe the
dynamics of the system it is sufficient to consider only the $k=0$ mode. 

Let us now consider the number of particles transported after a full period $T$. The
transport  can be characterized in terms of the integrated particle
flux, e.g.~between the $n$th and $n+1$st unit cell
\begin{equation}
\Phi_{n}=  i\int_{0}^{T} \!\! dt \, w_{2}\left(  t\right)  \left\langle \Psi\left(
t\right)  \right\vert \left(  \hat a_{n+1} \hat b_{n}^{\dagger}-\hat a_{n+1}^{\dagger}%
\hat b_{n}\right)  \left\vert \Psi\left(  t\right)  \right\rangle .\label{flux_beta}%
\end{equation}
Due to the translational symmetry of the  flux $\Phi_{n}$ does not depend on
$n$. Assuming that the initial amplitudes $\alpha$ and $\beta$ 
coincide with an eigenstate of the Hamiltonian $\mathbf{h}_{0}$, and slowly
varying the Hamiltonian parameters in time compared to the inverse energy gap
$1/(2 \varepsilon_{k=0}(t))$, leads to an adiabatic following of the many-body state.
Making use of the adiabatic approximation, after a straightforward
calculation, we find for the integrated particle flux
\begin{equation}
\Phi_{n}=\frac{1}{2}
{\displaystyle\oint_{\mathcal{C}}}
\frac{w_{1}+w_{2}}{\Delta^2\sqrt{\Delta^{2}+\left(  w_{1}+w_{2}\right)  ^{2}}%
}\bigl(  \Delta dw_{2}-w_{2}d\Delta\bigr) 
,\label{flux_geometric}
\end{equation}
where $\mathcal{C}$ is a closed path in the parameter space $\left(
w_{1},w_{2},\Delta\right)$. One recognizes that the flux $\Phi_{n}$ can also be evaluated using
Stokes' theorem by expressing it as an integral
of a vector $\mathbf{B}=( w_{1},w_{2},\Delta)/\bigl((w_1+w_2)^2+\Delta^2\bigr)^{3/2}$ through an area element $d\mathbf{S}$ in this
parameter space $\Phi_{n}=\frac{1}{2}\int_{\mathcal{S}}
\mathbf{B\cdot}d\mathbf{S}$,
where $\mathcal{S}$ is a surface with boundary $\mathcal{C}$. Hence, after an integer number of cycles there is a
net particle geometric transport which is however not quantized (topological).

To be specific, we have shown in Fig.~(\ref{fig:flux}) the integrated particle
current as function of the rescaled cycle time $AT$ with hopping rates
$w_{1}\left( t\right)  =A\cos^{2}\left(  \frac{\pi t}{T}\right), w_{2}\left(  t\right)  =A\sin^{2}\left(  \frac{\pi t}{T}\right) $ and
$\Delta\left(  t\right)  =A\sin\left(  \frac{2\pi t}{T}\right) $. The
horizontal dashed line shows the adiabatic value
\begin{align}
	\Phi_{n} = \frac{1}{2}
	{\displaystyle\int_{0}^{\pi}}
		\frac{\cos^{2}\left(  t\right)  }{\left(  \sin^{2}\left(  t\right)  +1\right)
		^{3/2}}dt=\frac{\Gamma\left(  \frac{3}{4}\right)  }{\sqrt{2\pi}} \approx 0.6,
\end{align}
of the net particle transport.

While the particle transport is in general not quantized, the polarization (\ref{eq:Resta}) can only change by an integer valued amount 
upon a full cycle of evolution, since it is the phase of a complex function (modulo $2\pi$), provided there are no transitions to other states. The latter is guaranteed by the adiabatic evolution.
In the above case one finds that the polarization winding of the bosonic Rice-Mele model vanishes. In fact one can easily calculate the polarization at any time $t$ exactly.
Fixing the gauge, i.e.~fixing the origin of the spatial coordinate on the circle of length $L$, one obtains
\begin{equation}
P=\frac{1}{2\pi}\arg\Bigl[  \exp\Bigl\{  -L\left(\vert \alpha(t)\vert^2 +\vert\beta(t)\vert^2\right) \Bigr\} \Bigr]=0, \label{PP}
\end{equation}
where we have evaluated the unitary operator $\hat T$ using its normally ordered form
\begin{equation}
\! \hat T \!  \, \, =:\prod_{r,s}\exp\left\{ \Bigl(e^{\frac{2\pi i}{L} (r + s/n)}-1\Bigr)
\hat a_{r,s}^{\dagger} \hat a_{r,s}\right\}: \, . \label{Polarization-normal}%
\end{equation}
The polarization is therefore constant in time. Clearly, there is no connection between the net particle transport and the change of the many-body polarization.
But it is even more surprising that the latter does not wind irrespective of the path taken in parameter space.
We will show in the following that the absence of polarization winding is a generic feature of Gaussian bosonic systems which is in sharp contrast to the fermionic analogue. 

%%%%%%%%%%%%%%%%%%%%%%%%%%%%%%
\section{polarization for bosons}
%%%%%%%%%%%%%%%%%%%%%%%%%%%%%%

The goal of this section is to calculate the expectation value of the unitary operator 
\begin{equation}
\hat T=\exp\left(  \frac{2\pi i}{L}{\displaystyle\sum\limits_{r,s}}
\left(r+\frac{s}{n}\right) \, \hat a_{r,s}^{\dagger} \hat a_{r,s}\right). \label{Polarization}%
\end{equation}
Here $\hat a_{r,s}^{\dagger}, \hat a_{r,s}$ are bosonic creation and annihilation operators respectively, where $r=0,\dots,L-1$ labels unit cells and $s=0,\dots,n-1$ internal sites in the unit cell.
$0\le \frac{s}{n} < 1$ and we have set the lattice constant equal to unity. 
The results of the following discussion do also not depend on the dimension of the system nor the total number of particles.
We note that the operator $\hat T$ is not gauge invariant because it changes under an arbitrary shift of the origin of the spatial coordinate system.
Throughout this paper we choose a coordinate system in which $\exp\left(  \frac{2\pi i}{L}(r+\frac{s}{n})\right) \neq 1$ for any $r,s$.

We consider a general bosonic Gaussian state \cite{gaussian,Holevo-PRA-1999} $\rho$ which can be formally expressed in diagonal form (Glauber-Sudarshan representation \cite{Sudarshan-PRL-1963,Glauber-PRL-1963}) in terms of multi-mode coherent states
\begin{equation}
\rho=\int d^{2}\vec{\mathbf \alpha}\, {\cal P}\left(
\vec{\mathbf{\alpha}}\right) \,  \left\vert
\vec{\mathbf{\alpha}}\right\rangle \left\langle
\vec{\mathbf{\alpha}}\right\vert , \label{density11}%
\end{equation}
where $d^2 \alpha = d\alpha_r d\alpha_i$, with $\alpha_r=(\alpha+\alpha^*)/2$ and $\alpha_i=(\alpha-\alpha^*)/(2i)$ being the real and imaginary parts of the coherent amplitude
\begin{align}
{\cal P}\left(  \vec{\alpha}\right) = \mathcal{N}{\displaystyle\int} d^{2}\vec{\mathbf{\eta}}\exp\bigg( &-\frac{1}{2} \vec{\mathbf{\eta}}^{T}\left(  \matr{V}-\id\right) \vec{\mathbf{\eta}}\nonumber\\
&-i\left(2\vec{\mathbf{\alpha}}+\vec{\alpha}_0\right)^{T}\vec{\mathbf{\eta}}\bigg). \label{p_function22}
\end{align}
Here ${{1\!\!1}}$, $\vec{\mathbf{\alpha}}=\bigl((\alpha_{1,r},\alpha_{1,i}),(\alpha_{2,r},\alpha_{2,i})\dots\bigr)$ and $\vec{\mathbf{\eta}}= \bigl((\eta_{1,r},\eta_{1,i}),(\eta_{2,r},\eta_{2,i})\dots\bigr )$ represent the identity matrix and real vectors respectively with dimension $2nL$ (note that $nL$ is the number of bosonic modes of the problem).
$\mathcal{N}$ is a normalization constant ensuring that $\int d^{2}\vec{\mathbf{\alpha}}\, {\cal P}\left(  \vec{\mathbf{\alpha}}\right) = 1$.
The explicit form of $\mathcal{N}$ is not relevant for our purposes. $\vec{\alpha}_0 = (\langle \a + \ad \rangle, -i \langle \a - \ad \rangle)^T$ encodes the expectation values of the mode operators and $\matr{V}$ is the $2nL\times 2nL$ covariance matrix of the system, which for a single mode and $n=1$ reads
\begin{equation}
\matr{V} =\left(\begin{array}{cc}
\langle\langle \hat q \hat q \rangle\rangle & \frac{1}{2} \langle\langle \hat p\hat q +\hat q\hat p\rangle\rangle\\
\frac{1}{2} \langle\langle \hat p\hat q +\hat q\hat p\rangle\rangle & \langle\langle \hat p\hat p\rangle\rangle
\end{array}
\right).
\end{equation}
Here $\hat q = \hat a+\hat a^\dagger$ and $\hat p=-i(\hat a -\hat a^\dagger)$, and $\langle\langle xy\rangle\rangle = \langle xy\rangle -\langle x\rangle\langle y\rangle$.
$\matr{V}$ is a real and symmetric matrix by construction and is also positive definite due to the Heisenberg uncertainty principle. ${\cal P}$ is positive and well defined if furthermore $\matr{V}>\id$. In this case the state is a statistical mixture of coherent states, i.e.~is a classical state.
A quantum state is considered to be nonclassical if it cannot be written as a statistical mixture of coherent states.
In this paper we consider more general bosonic Gaussian states (A good introduction to bosonic Gaussian states can be found, for example, in \cite{gaussian}).

$\cal{P}\left( \vec{\mathbf{\alpha}} \right)$ can be used to evaluate the expectation value of any normally ordered operator function $:\! f(\{\hat a^\dagger_\mu,\hat a_\mu\})\!:$ by the replacement $(\hat a^\dagger \to \alpha^*)$ and $(\hat a \to \alpha)$ and integration. 
The ${\cal P}$ function may be singular and can attain negative values.
All integration with ${\cal P}\left( \vec{\mathbf{\alpha}} \right)$ must therefore be understood in the distributional sense. 

Using eq.~(\ref{Polarization-normal}) we find
\begin{eqnarray}
\langle \hat T\rangle &=&\mathcal{N}_{1}\int \!\! d^2 \vec\eta\int\!\! d^{2} \vec\alpha\exp\Bigl\{  -\frac{1}{2}\vec{\mathbf{\eta}}^{T}\left(\matr{V}-\id\right)  \vec{\mathbf{\eta}}-i\vec{\mathbf{\alpha}}_0 ^{T}\vec{\mathbf{\eta}}\Bigr\}\nonumber\\
&& \quad \times\exp\Bigl\{-2i\vec{\mathbf{\alpha}}\vec{\mathbf{\eta}^{T}}-\vec{\mathbf{\alpha}}^{T}\left(\id-\matr{U}\right) \vec{\mathbf{\alpha}}\Bigr\}  , \label{Gaussian_Integral}%
\end{eqnarray}
where $\matr{U}$ is a unitary operator
\begin{equation}
\left(  \matr{U}\right)  _{r_1,s_1;r_2,s_2}=\exp\left(  \frac{2\pi i}{L} \left(r_1 + \frac{s_1}{n}\right)\right)  \delta_{r_1r_2}\delta_{s_1,s_2}.
\label{unitary}%
\end{equation}
According to our assumption ($\exp\left(  \frac{2\pi i}{L}(r + \frac{s}{n})\right)  \neq1$), ${{1\!\!1-}}\matr{U}$ is an invertible symmetric complex matrix.
In addition, its real part ${{1\!\!1-}}\frac{\matr{U}+\matr{U}^{\dagger}}{2}$ is positive definite.
In this case the Gaussian integral (\ref{Gaussian_Integral}) over $\vec{\mathbf{\alpha}}$ is well-defined and is proportional to $\left[  \det\left( {{1\!\!1-}}\matr{U} \right) \right]^{-1/2}$.
We note that when the matrix is complex, the calculation of the square root requires some special care. However, one can show that any symmetric complex matrix has a unique symmetric square root whose real part is positive definite \cite{Kato}.
After successive integration over $\vec{\mathbf{\alpha}}$ and then over $\vec{\mathbf{\eta}}$ we eventually obtain
\begin{eqnarray}
\langle \hat T\rangle &=&\mathcal{N}_{2}\left[ \det\left( \matr{V}+\id\right)  \det\left(  \id-\frac{\matr{V}-\id}{\matr{V}+\id}\matr{U}\right)
\right]  ^{-1/2}\nonumber\\
&&\qquad \times \exp\left(-\frac{1}{2}\vec{\mathbf{\alpha}}_0^T \matr{M^{-1}} \vec{\mathbf{\alpha}}_0\right),\label{Final_Expression}
\end{eqnarray}
where $\mathcal{N}_{2} = 2^{nL}$ and
\begin{align}
	\matr{M}=\matr{V}-\id+2(\id-\matr{U})^{-1}. \label{eq:M_matrix}
\end{align}
Substituting this expectation value into the expression of the many-body polarization (\ref{eq:Resta}) one obtains
\begin{eqnarray}
P&=&-\frac{1}{4\pi}\im \ln \left[  \det\left(  \matr{V}+\id\right)  \det\left(
\id-\frac{\matr{V}-\id}{\matr{V}+\id}\matr{U}\right)  \right] \nonumber \\
&& \qquad -\frac{1}{4\pi} \textrm{Im} \left(-\frac{1}{2}\vec{\mathbf{\alpha}}_0^T \matr{M^{-1}} \vec{\mathbf{\alpha}}_0\right)\label{Final_Exp}\\
&=&-\frac{1}{4\pi}\operatorname{Im}\ln\Bigl[  \det\bigl(  {{1\!\!1-\matr{W}}}\bigr)
\Bigr]  -\frac{1}{4\pi} \textrm{Im} \left(-\frac{1}{2}\vec{\mathbf{\alpha}}_0^T \matr{M^{-1}} \vec{\mathbf{\alpha}}_0\right) ,\nonumber %
\end{eqnarray}
where
\begin{equation}
\matr{W}=\frac{{{\matr{V}-\id}}}{{{\matr{V}+\id}}}\matr{U}. \label{WWW}%
\end{equation}
One can show that the second term in eq.~(\ref{Final_Exp}) is a single valued function of system parameters and therefore does not contribute to the change of polarization.
In the next section we will show that the first term in eq.~(\ref{Final_Exp}) vanishes in the thermodynamic limit of infinite system size $L\to\infty$.

%%%%%%%%%%%%%%%%%%%%%%%%%%%%%%%%%%
\section{Polarization in the thermodynamic limit}
%%%%%%%%%%%%%%%%%%%%%%%%%%%%%%%%%%

%%%%%%%%%%%%%%%%%%%%%%%%%%%%%%%%%%
\subsection{Polarization scaling: bosons vs. fermions}
%%%%%%%%%%%%%%%%%%%%%%%%%%%%%%%%%%

In Ref.~\cite{Bardyn-PRX-2018} it was shown that the polarization of a general Gaussian mixed state $\rho$ of lattice fermions at commensurate filling can be written as a sum of the polarization of a pure state $\vert \psi\rangle$ plus a term that vanishes in the thermodynamic limit of infinite system size $L\to\infty$. 
\begin{equation}
P(\rho) = P\bigl(\vert \psi\rangle\langle \psi\vert\bigr) + {\cal O}(L^{-\alpha}),\quad \alpha >0.
\end{equation}
Here $\vert\psi\rangle$ is the many-body ground state of the so-called fictitious Hamiltonian. 
In the following we will assume that the second term in eq.~(\ref{Final_Exp}) vanishes and show that the remaining term in the bosonic case yields
\begin{equation}
P(\rho) = 0 + {\cal O}(e^{-\alpha L}),\quad \alpha > 0.\label{eq:P-bosonic}
\end{equation}
For simplicity we restrict ourselves to the simplest non-trivial case of a two-band model, e.g.~resulting from a tight-binding Hamiltonian with a unit cell of two lattice sites. The generalization to the case of multiple bands is however straight forward.

We introduce the Fourier transform given by the unitary block matrix $\matr{U}_{FT}$
\begin{align}
	(\matr{U}_{FT})_{jk} \equiv \frac{1}{\sqrt{L}} \exp\left(\frac{2\pi i}{L} jk\right) 1\!\!1_4.
\end{align}
As a consequence of the periodic boundary conditions the covariance matrix $\matr{V}$ is block-circulant. Since the model has lattice translational invariance, the covariance matrix is diagonalized by the Fourier transform and we can write:
\begin{align}
	\matr{U}_{FT} \frac{\matr{V}-\id_{4L}}{\matr{V}+\id_{4L}} \matr{U}^\dagger_{FT} = \bigoplus_{k=0}^{L-1} \frac{\matr{v_k}-\id_{4}}{\matr{v_k}+\id_{4}},
\end{align}
where $\oplus$ denotes the direct sum which constructs a block diagonal matrix.
The transformed unitary matrix $\matr{U}$, given by eq.~(\ref{unitary}), is:
\begin{align}
	(\matr{U}_{FT} \, \matr{U} \, \matr{U}_{FT}^\dagger)_{j,s_1;k,s_2} &= \delta_{j,k+1} \delta_{s_1 s_2} \exp\left(\frac{2\pi i}{L} \frac{s_1}{n} \right) \id_{4}.
\end{align}
To make the following expressions more compact, we furthermore introduce $\matr{m}_k \equiv \frac{\matr{v_k}-\id_4}{\matr{v_k}+\id_4} (\id_2 \oplus \exp\left(\frac{i\pi}{L}\right) \id_2)$.
The determinant in eq.~(\ref{Final_Exp}) can thus be written as
\begin{align}
	&\det \left( \id - \frac{\matr{V}-\id}{\matr{V}+\id} \matr{U} \right) = \\
	&\qquad =
	\begin{vmatrix}
		\id_{4} & 0 & & & & -\matr{m}_{L-1}\\
		-\matr{m}_0 & \id_{4} & 0 \\
		0 & -\matr{m}_1 & \id_{4} & 0 \\
		& \ddots & \ddots & \ddots & \ddots\\
		& & 0 & -\matr{m}_{L-3} & \id_{4} & 0\\
		0 & & & 0 & -\matr{m}_{L-2} & \id_{4}
	\end{vmatrix}.\nonumber
\end{align}
This block determinant can be reduced by applying Schur's identity iteratively. This yields a determinant of dimension $4\times 4$:
\begin{align}
	\det \left( \id - \frac{\matr{V}-\id}{\matr{V}+\id} \matr{U} \right) &= \det\left( \id_4 - \prod_{k=0}^{L-1} \matr{m}_k \right). \label{gauge_reduction}
\end{align}
We note that up to this point there is a formal analogy of the polarization for Gaussian states of bosons and that of fermions, discussed in Ref.~\cite{Bardyn-PRX-2018}.
There the matrices $\matr{m}_k\sim e^{-B_k} U_{k+1}^\dagger U_k$ contained unitary matrices $U_k$ and weighting factors $e^{-B_k}=\textrm{diag}_s\bigl(e^{-\beta_{k,s}}\bigr)$. To be specific let us consider a grand-canonical thermal state of a fermionic insulator with a chemical potential $\mu$ within a band gap. Then all bands $s$ with energies below $\mu$ lead to a negative exponent $\beta_{k,s} = \beta (\epsilon_{k,s} -\mu)$ and thus to weighting factors bigger than unity. This results in an amplification of contributions from occupied bands, which is the essence of the gauge-reduction mechanism for Gaussian states of fermions found in \cite{Bardyn-PRX-2018}. 

The situation is completely different, however, in the case of bosons. 
Since the covariance matrix $\matr{V}$ of Gaussian states of bosons is positive definite, the resulting k-dependent $4\times 4$ blocks have eigenvalues $\lambda\left(\matr{m}_k\right)$ with absolute values obeying
\begin{align}
	\left|\lambda\left(\matr{m}_k\right)\right| &< 1 \qquad \forall k=0,\dots,L-1.
\end{align}
We define the corresponding maximum absolute eigenvalue:
\begin{align}
	\lambda_\text{max} &\equiv \max_i \left|\lambda_i\left(\frac{\matr{V} - \id}{\matr{V} + \id}\right)\right|,\label{eq:lambdas_m}
\end{align}
%
% and minimal absolute eigenvalues $\lambda_\text{min}^\text{V}$, $\lambda_\text{min}^\text{m}$ accordingly.
% These definitions are summarized by:
%\begin{align}
%	0 < \lambda_\text{min}^\text{V} &\leq \lambda_\text{max}^\text{V} < \infty,\\
%	0 < \lambda_\text{min}^\text{m} &\leq \lambda_\text{max}^\text{m} < 1.
% \end{align}
%
According to eq.~(\ref{eq:lambdas_m}), a single matrix $\matr{m}_k$ is bounded and thus the product of matrices must be bounded as well, i.e.~$\left\Vert \prod_k \matr{m}_k \right\Vert = \mathcal{O}((\lambda_\text{max})^L)$.
If $\lambda_\text{max} = 0$ the polarization vanishes trivially, else we split off the maximum absolute eigenvalues $\matr{A} \equiv (\lambda_\text{max})^{-L} \prod_k \matr{m}_k$ such that $|\tr (\matr{A})| \leq 4$ and define a small parameter $\epsilon \equiv 4 (\lambda_\text{max})^{L}$.
We can then express the polarization $P$ by expanding the determinant and logarithm in this small parameter:
\begin{align}
	\ln\det\left( \id_{4} - \prod_k \matr{m}_k \right) =& \ln\det \left( \id_{4} - \frac{\epsilon}{4} \matr{A} \right) \nonumber\\
	=& \ln \left( 1 - \frac{\epsilon}{4} \tr (\matr{A}) + \mathcal{O}(\epsilon^2) \right) \nonumber\\
	=& - \frac{\epsilon}{4} \tr (\matr{A}) + \mathcal{O}(\epsilon^2).
\end{align}
With this we find the following system-size scaling of the polarization for Gaussian bosonic states
\begin{align}
	4\pi|P| \leq& \frac{\epsilon}{4}|\tr(\matr{A})| + \mathcal{O}(\epsilon^2) \nonumber\\
	\leq& \epsilon + \mathcal{O}(\epsilon^2).
\end{align}
Since we know that $0 \leq \lambda_\text{max} < 1$, the small parameter $\epsilon$ vanishes exponentially in $L$,
\begin{align*}
\alpha \equiv -\ln(\lambda_\text{max}) > 0 \implies \epsilon = 4 e^{-\alpha L}.
\end{align*}
Therefore, as the system approaches the thermodynamic limit, the first term of the many-body polarization in eq.~(\ref{Final_Exp}) vanishes exponentially and only the trivial second term remains. 
For equilibrium states at finite $T$ this has a simple physical interpretation: The chemical potential for (non-interacting) bosons is always less than the smallest single-particle energy.
As a consequence \emph{all} weighting factors $e^{-\beta_{k,s}}$ are strictly less than unity and there is no amplification that leads to a gauge reduction as in the case of fermions.
Thus the absence of the Pauli exclusion principle for (non-interacting) bosons also leads to the absence of a gauge reduction mechanism as in Ref.~\cite{Bardyn-PRX-2018}.

As an illustration of our results, we analyze the bosonic Rice-Mele model with the initial state (\ref{eq:coherent}). The covariance matrix of this state is just the identity \cite{gaussian} and therefore, as was expected the expression of $P$ eq.~(\ref{Final_Exp}) coincides with eq.~(\ref{PP}).

%%%%%%%%%%%%%%%%%%%%%%%%%%%%%%%%%%
\subsection{Polarization amplitude}
%%%%%%%%%%%%%%%%%%%%%%%%%%%%%%%%%%

Since the many-body polarization is defined as the complex phase of the lattice momentum shift, it can only be defined if the absolute value $\vert \langle \hat{T} \rangle\vert $ does not vanish throughout the entire adiabatic evolution.
It turns out that this is always true for finite system sizes. However, as noted by Resta and Sorella $\vert \langle \hat{T} \rangle\vert $ is a measure for the localization of single-particle states \cite{Resta-Sorella-PRL-1999}, which in the thermodynamic limit approaches unity for an insulator and vanishes for a conductor. Thus 
for non-interacting bosons
we expect it to decay when $L\to \infty$. Both can be seen by inserting eq.~(\ref{gauge_reduction}) into eq.~(\ref{Final_Expression}) and taking the absolute value:
\begin{eqnarray}
	&&  |\langle \hat{T}\rangle | = 2^{nL} |\det(\id_{2nL} + \matr{V})|^{-1/2} \label{eq:T_abs}\\
	&& \cdot \left|\det\left(\id_{2n} - \prod_{k=0}^{L-1}\matr{m}_k\right)\right|^{-1/2}\, \left|\exp\left( -\frac{1}{2} \vec{\alpha}_0^T \matr{M}^{-1} \vec{\alpha}_0 \right)\right|.  \nonumber
\end{eqnarray}
We proceed by finding upper and lower bounds. To this end we note that for the absolute value of the last exponential term only the hermitian part of the matrix contributes
$\frac{1}{2}(\matr{M} + \matr{M}^\dagger) = \matr{V}$. Thus 
\begin{align}
	0 < \left|\exp\left( -\frac{1}{2} \vec{\alpha}_0^T \matr{M}^{-1} \vec{\alpha}_0 \right)\right| < 1. \label{eq:exponential_bound}
\end{align}
From this one can see that $\vert \langle\hat T\rangle\vert$ is always positive for finite system sizes $L$.
Denoting the minimum eigenvalue of $\matr{V}$ by $\lambda_\text{min}^\text{V}$ and assuming a classical state, i.e. $\lambda_\text{min}^\text{V} > 1$, one can derive an upper bound which scales in the system size $nL$:
\begin{align}
	0 < |\langle \hat{T}\rangle | < \left(\frac{1+\lambda_\text{min}^\text{V}}{2}\right)^{-nL} < 1.
\end{align}
From this we can see that the many-body polarization $P$ is well-defined for all system sizes $L<\infty$ and that classical states exhibit negligible single-particle localization for large $L$.

%%%%%%%%%%%%%%%%%%%%%%%%%%%%%%%%%%
\section{Polarization winding}
%%%%%%%%%%%%%%%%%%%%%%%%%%%%%%%%%%

In one-dimensional
lattice systems with a Hamiltonian or a Liouvilian which depend on an external parameter $\lambda$
in a cyclic way, the winding of the EGP or the many-body polarization with $\lambda$
defines a topological invariant:
\begin{equation}
w=\Delta P = \oint d\lambda \frac{\partial P(\lambda)}{\partial \lambda}
\end{equation}
In two-dimensional translational invariant lattice models a similar construction defines a Chern number.
E.g.~introducing particle number operators in mixed real and momentum space by performing a discrete Fourier-transformation
in one direction, (e.g.~$y$), $\hat a_j(k_y)\sim \sum_l \hat a_{j,l} \exp(2\pi i l k_y/L)$, one can define a momentum-dependent polarization
(where we have suppressed band indices for simplicity)
\begin{equation}
P_x(k_y) = \frac{1}{2\pi} \textrm{Im}\ln \left\langle  \exp\left(\frac{2\pi i}{L}\sum_j j \hat a_j^\dagger(k_y) \hat a_j(k_y)\right)\right\rangle.
\end{equation}
The winding of $P(k)$ when going through the Brillouin zone in $k$, defines a Chern number
\begin{equation}
C = \int_\textrm{BZ} dk_y \frac{\partial P_x(k_y)}{\partial k_y} =  \int_\textrm{BZ} dk_x \frac{\partial P_y(k_x)}{\partial k_x}
\end{equation}
If we consider the polarization in a Gaussian mixed state of bosons $\rho(\lambda)$, which
is uniquely defined along a closed path of the parameter $\lambda$ in parameter space,
we can argue from eq.~(\ref{eq:P-bosonic}) that the winding of the many-body polarization to vanish for a sufficiently large but finite system size $L$. This is because of the exponential bound that yields $-\frac{1}{2} < P < +\frac{1}{2}$ if the system is large enough.
As a consequence all many-body topological invariants based on the winding of the polarization
are trivial for sufficiently large 
% but finite and infinite 
systems.
In the following we will explicitly show that this holds true independently of the system size.

Let us assume that the polarization is a function of two real parameters which change cyclically in time from $0$ to $T$. Then the change of the polarization between times $t=0$ and $t=T$ can be described as a loop
along a closed path $\mathcal{C}$ in parametric space.
The two parameters can be combined into a complex variable $\chi$. Thus the change of polarization can be written as
\begin{equation}
\Delta P=-\frac{1}{4\pi}\operatorname{Im}
{\displaystyle\oint\limits_{\mathcal{C}}}
d\chi\frac{\partial}{\partial \chi}\ln\det\bigl[  \id_{2nL}-\matr{W}\left(  \chi\right)  \bigr].
\label{change_Polarization}
\end{equation}
Moreover, using
\begin{equation}
\frac{\partial}{\partial\chi}\ln\det\left[  {{1\!\!1-}}\matr{W}\left(  \chi\right)  \right]
=\textrm{Tr}\left[  \left[  {{1\!\!1-}}\matr{W}\left(  \chi\right)  \right]
^{-1}\frac{\partial (\id-\matr{W}(\chi))  }{\partial\chi}\right]  .
\label{derivativedeterminant}
\end{equation}
we derive the following expression for $\Delta P$
\begin{equation}
\Delta P=\frac{1}{4\pi}\operatorname{Im}\textrm{Tr}
{\displaystyle\oint\limits_{\mathcal{C}}}
d\chi\left[  \left[  1\!\!1-\matr{W}\left(  \chi\right)  \right]  ^{-1}
\frac{\partial (\id - \matr{W}(\chi))  }{\partial \chi}\right]  . \label{Simple_Form}
\end{equation}
The expression (\ref{derivativedeterminant}) can be derived from the identity $\ln(\det(A(z))=\textrm{Tr}(\ln A(z))$ (for a rigorous derivation of (\ref{derivativedeterminant}) the reader is referred to \cite{Krein}.

Now we are ready to prove that the change of polarization vanishes for any
bosonic Gaussian state. For that we first review some facts about zeros of
determinants of holomorphic matrix-valued functions (for more details see
\cite{Gohberg}).

Let $\matr{F}\left(  \chi\right)  $ be a matrix-valued function that is analytic in a
domain $C$. Under the assumption that all values of $\matr{F}\left(  \chi\right)  $
on the boundary $\mathcal{C}$ of $C$ are invertible operators it is possible to
show \cite{Gohberg} that
\[
{\cal M}=\frac{1}{2\pi i}\textrm{Tr}{\displaystyle\oint\limits_{\mathcal{C}}}
d\chi\left[  \matr{F}\left(  \chi\right)  ^{-1}\frac{d\matr{F}\left(  \chi\right)  }{d\chi
}\right]
\]
is the number of zeros of $\det \matr{F}\left(  \chi\right)$ inside $\mathcal{C}$ (including their multiplicities). Combining this with equation (\ref{Simple_Form}),
we obtain
\begin{equation}
\Delta P=\frac{1}{2}{\cal M}, \label{zeros}
\end{equation}
where ${\cal M}$ is the number of solutions (zeros) of
\[
\det\left[  1\!\!1-\matr{W}\left(  \chi\right)  \right]  =0
\]
inside the closed path $\mathcal{C}$ in parametric space. In order to estimate
${\cal M}$, we use a generalization of Rouch\'e's theorem for the
matrix valued complex function \cite{Gohberg}, which states:

\textit{Rouch\'e's Theorem}: \ Let $\mathcal{C}$ be a closed contour bounding a
domain $C$. If $\left\Vert \matr{F}\left(  \chi\right)  \right\Vert <1$ on
$\mathcal{C}$ then
\[
\frac{1}{2\pi i}\textrm{Tr}
{\displaystyle\oint\limits_{\mathcal{C}}}
d\chi\left[  \left(  {{1\!\!1+}}\matr{F}\left(  \chi\right)  \right)  ^{-1}
\frac{d \matr{F}\left(  \chi\right)  }{d\chi}\right]  =0.
\]
Applying Rouch\'e's theorem to our problem, where
\[
\left\Vert \matr{F}\left(  \chi\right)  \right\Vert =\left\Vert \matr{W}\left(  \chi\right)
\right\Vert =\left\Vert \frac{\matr{V}-1\!\!1}{\matr{V}+1\!\!1}\right\Vert <1
\]
we see that for any $\matr{V}>0$, i.e.~for any Gaussian bosonic state
\[
\left\Vert \matr{W}\left(  \chi\right)  \right\Vert <1.
\]
Therefore the change of polarization is equal to zero, irrespective of the
system size.
\begin{equation}
\Delta P = 0.
\end{equation}
We note that this result is again a direct consequence of the positivity of the covariance matrix $\matr{V}$ for Gaussian states
of \emph{bosons}.
This proves that for any bosonic Gaussian state the total change of the many-body
polarization along a closed path in parametric space is zero. This is in sharp contrast to free
fermion systems in which the winding of the many-body polarization is a
topologically quantized observable and can be non-trivial.

%%%%%%%%%%%%%%%%%%%%%%%%%%%%%%%%%%
\section{conclusion}
%%%%%%%%%%%%%%%%%%%%%%%%%%%%%%%%%%

We have shown that the many-body polarization of translationally invariant Gaussian states of bosons approaches zero in the 
thermodynamic limit of infinite system size. Its winding upon a cyclic change of the state, which in the case of fermions defines a many-body topological invariant, vanishes
for any system size. Thus many-body topological invariants based on the polarization are always trivial in finite-temperature states or Gaussian non-equilibrium states of non-interacting bosons. This is also the case if the band structure of the underlying lattice Hamiltonian is topologically non-trivial, i.e.~possesses bands with a non-vanishing Chern 
number. As a consequence there is no topologically protected quantized charge transport of Gaussian states of bosons and the latter requires strong interactions
\cite{Lohse-NatPhys-2015}. This property of bosons is in sharp contrast to fermions, which can be topologically non-trivial even in
many-body states that are not gapped, such as high-temperature states of band insulators, 
 and is a consequence of the absence of a Pauli exclusion principle.

%%%%%%%%%%%%%%%%%%%%%%%%%%%%%%%%%%
\section*{Acknowledgement}
Financial support from the DFG (project number 277625399) through SFB TR 185 is gratefully acknowledged. 

%%%%%%%%%%%%%%%%%%%%%%%%%%%%%%%%%%%%%%%%%%%%%%%%%%%%%%%%%%%%%%%%%

\end{document}